\documentclass[aps,reprint,prl,superscriptaddress,floats,floatfix]{revtex4-2}

\usepackage{amsmath}
\usepackage{amsfonts}
\usepackage{amssymb}
\usepackage{graphicx}
\usepackage{color}
\usepackage{tikz}
\usepackage{lipsum}
\usepackage{float}
\usepackage[colorlinks=true,linkcolor=blue,urlcolor=blue,citecolor=blue,pdfusetitle]{hyperref}
\usepackage{amstext}
\usepackage{latexsym}
\usepackage[running,mathlines]{lineno}
\usepackage{bbm}
\usepackage{verbatim}

\usepackage{ulem}

\definecolor{dgreen}{rgb}{0.0, 0.5, 0.0}
\definecolor{rot}{RGB}{165,30,55}

\newcommand{\bc}[1]{{\color{blue}{#1}}}

\newcommand{\st}[1]{{\color{blue}{#1}}}

\newcommand{\df}{\mathrm{d}}
\newcommand{\Tr}{\mathrm{Tr}}

\newcommand{\tp}{\text{ . }}
\newcommand{\tc}{\text{ , }}

\renewcommand{\em}{\itshape}
\renewcommand{\emph}[1]{\textit{#1}}

\newcommand{\tubingen}{Institut f\"ur Theoretische Physik, Universit\"at Tübingen and Center for Integrated Quantum Science and Technology, Auf der Morgenstelle 14, 72076 T\"ubingen, Germany}

\newcommand{\Nottingham}{School of Physics and Astronomy and Centre for the Mathematics and Theoretical Physics of Quantum Non-Equilibrium Systems, The University of Nottingham, Nottingham, NG7 2RD, United Kingdom}

\newcommand{\Coventry}{Centre for Fluid and Complex Systems, Coventry University, Coventry, CV1 2TT, United Kingdom}

\newcommand{\Mallorca}{Institute for Cross-Disciplinary Physics and Complex Systems (IFISC) (UIB-CSIC), E-07122 Palma de Mallorca, Spain}

\newcommand{\TubingenMath}{Mathematisches Institut, Eberhard-Karls-Universität, Auf der Morgenstelle 10, 72076 Tübingen,
Germany}

\begin{document}
	
\title{Adiabatically driven dissipative many-body quantum spin systems}

\author{Paulo J. Paulino}
\email{paulo@ifisc.uib-cisc.es}
\affiliation{\tubingen}
\affiliation{\Mallorca}
\author{Stefan Teufel}
\affiliation{\TubingenMath}
\author{Federico Carollo}
\affiliation{\Coventry}
\author{Igor Lesanovsky}
\affiliation{\tubingen}
\affiliation{\Nottingham}

\date{\today}

\begin{abstract}    
    We explore the evolution of a strongly interacting dissipative quantum Ising spin chain that is driven by a slowly varying time-dependent transverse field. This system possesses an extensive number of instantaneous (adiabatic) stationary states which are coupled through non-adiabatic transitions. We analytically calculate the generator of the ensuing slow dynamics and analyze the creation of coherences through non-adiabatic processes. For a certain choice of the transverse field shape, we show that the system solely undergoes transitions among classical basis states after each pulse. The concatenation of many of such pulses leads to an evolution of the spin chain under a many-body dynamics that features kinetic constraints. Our setting not only allows for a quantitative investigation of adiabatic theorems and non-adiabatic corrections in a many-body scenario. It also directly connects to many-body systems in the focus of current research, such  as ensembles of interacting Rydberg atoms which are resonantly excited by a slowly varying laser pulse and subject to dephasing noise. 
\end{abstract}

\maketitle

\textit{Introduction.---} Adiabatic dynamics are relevant for many applications. These include adiabatic quantum computation~\cite{RevModPhys.90.015002,aharonov2008adiabatic,PhysRevA.65.042308,NP_ising_Andrew_2014}, protocols for the coherent manipulation of matter, such as stimulated Raman adiabatic passage~\cite{RevModPhys.89.015006}, and quantum devices~\cite{PRXQuantum.3.020347,Hu_2022,PhysRevE.100.032107,PhysRevA.103.062215,PhysRevA.91.022309,PhysRevX.3.021015,PhysRevLett.114.113901,PhysRevA.76.062304,RevModPhys.91.045001}. 
In adiabatically evolving systems the dynamics takes place within eigenspaces of a time-dependent Hamiltonian,  that are separated by a spectral gap and thus decoupled at leading order. Such an adiabatic limit has also been explored for open quantum systems,  whose evolution is governed by the so-called Lindblad master equation~\cite{breuer2002theory, sarandy2005adiabatic,avron2012}, where it entails the decoupled evolution of the eigenmatrices of the Lindbladian. In contrast to closed quantum systems, dissipative dynamics bring  the system to a stationary state, which adds an additional timescale to the adiabatic dynamics~\cite{albash2012quantum,sarandy2005adiabatic,yi2007adiabatic,PhysRevLett.95.250503,PhysRevA.93.032118}. Open quantum systems  evolving adiabatically are  relevant in a number of setups, including optimal control~\cite{Alipour2020shortcutsto, yin2022shortcuts, PhysRevA.104.062421,PhysRevA.103.012206}, adiabatic pumping~\cite{PhysRevB.92.245409}, noisy quantum computation~\cite{PhysRevLett.100.060503, PhysRevA.99.062320,PhysRevB.98.064307,PhysRevB.96.054301}, and the manipulation of stochastic emission records, i.e.\ quantum jump trajectories~\cite{PhysRevA.110.L060601,King2024adiabaticquantum,PhysRevLett.132.260402}. However, the complexity of describing the spectral properties of the Lindbladian, which is the generator of open quantum dynamics, makes it challenging to perform analytical studies of many-body scenarios \cite{sarandy2005adiabatic,avron2012}. Therefore, having an analytically treatable system would allow to gain quantitative insights into the structure of the generator of the adiabatic evolution and the structure and scaling of non-adiabatic corrections. 

\begin{figure}[h!]
    \centering
    \includegraphics[width=\linewidth,height=\textheight,keepaspectratio]{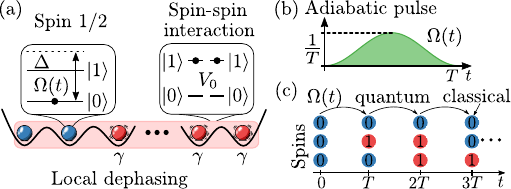}
    \caption{\textbf{Adiabatically evolving open spin chain.} We consider a chain formed by effective spin-$1/2$ particles (internal states $\left|0\right>$, $\left|1\right>$), e.g., representing two-level atoms. Transitions between the spin states are driven by a time-dependent transverse field (or laser) of strength $\Omega(t)$. An additional longitudinal field of strength $\Delta$ can be considered, which could be realized by detuning the laser with respect to the atomic resonance. Spins in the state $\left|1\right>$ interact (here with the nearest neighbor interaction strength $V_0$). The spins are subject to local dephasing at a rate $\gamma$. (b) Temporal profile of the transverse field $\Omega(t)$. We choose it to be a pulse of length $T$, with $\Omega(0) = \Omega(T) = 0$, and described by a smooth function. The pulse height is chosen $\propto T^{-1}$, so that its area is independent of $T$. (c) Concatenating several pulses  generates an effectively classical  kinetically constrained spin dynamics at the stroboscopic times $mT$, with $m$ being an integer number. Within each  interval $mT < t < (m+1)T$, however, quantum coherences are generated. The classical basis states $\left|0\right>$ and $\left|1\right>$ are represented by red and blue dots, respectively.} 
    \label{fig:Fig1}
\end{figure}

In this work, we study an adiabatically driven open quantum system, whose dynamics can to a large extent be studied analytically. It is formed by an Ising-type spin chain with power-law interactions, subject to strong local dephasing [see  Fig.~\ref{fig:Fig1}\bc{(a)}]. Adiabatic driving is implementing by applying a pulsed transverse field, as sketched in Fig.~\ref{fig:Fig1}\bc{(b)}. This setting is representative of a whole class of much-studied many-body models, instances of which can be experimentally realized, e.g., using Rydberg atoms \cite{Browaeys2020} or trapped ions \cite{yao2025}. On these platforms the spin degrees of freedom are represented by appropriately chosen electronic levels and the adiabatically varying transverse field is implemented by a slowly-varying laser pulse. To study the ensuing dynamics, we move into a rotating frame, where the instantaneous time-dependent eigenvalues and eigenmatrices of the Lindbladian generator can be calculated analytically. Following Ref.~\cite{avron2012} this enables the computation of the generator of the adiabatic dynamics and the first-order correction to it. For a particular choice of the transverse field profile the evolution generated by an adiabatic pulse gives rise to a quantum map that merely connects classical spin configurations. In this regime, the concatenation of multiple pulses [see  Fig.~\ref{fig:Fig1}\bc{(c)}] gives rise to a classical dynamics where the spins evolve under so-called kinetic constraints, that emerge through strong spin-spin interactions ~\cite{PhysRevLett.98.195702,Cancrini2008,lesanovsky2011,marcuzzi2014,PhysRevLett.126.103002,Cech_Kinetically_2025,Cech_Kinetically_2025,PhysRevE.102.062107}. 

\vspace{0.25cm}
\textit{Model.---} We consider a system of $N$ spins, each one characterized by the internal states $\left|0\right>$ and $\left|1\right>$, undergoing an open quantum evolution. The complete state of the system is encoded in the density matrix $\rho$ whose dynamics is governed by the Markovian quantum master equation 
\begin{eqnarray}\label{eq:master_equation}
    \frac{\partial}{\partial t} \rho = -i[H(t),\rho] + \sum_k \mathcal{D}_k[\rho]\, .\label{eq:initial_master_equation}
\end{eqnarray}
The operator $H(t)$ is a  time-dependent Hamiltonian  and assumes the form 
\begin{eqnarray}
    H(t) = \sum_{k=1}^N \left[ \Delta n_k + \Omega(t) \sigma^x_k\right] +  \frac{1}{2}\sum_{k,m=1}^N V_{km}n_k n_m\, ,
\end{eqnarray}
where $\sigma_k^\alpha$ is a Pauli matrix ($\alpha=x,y,z$) and $n_k=(1+\sigma^z_k)/2$ is the  projector onto the $\left|1\right>$-state, both acting on the $k$-th spin. In an experimental realization of this system with atoms or ions, the transverse field $\Omega(t)$ represents a time-dependent laser Rabi frequency and the longitudinal field $\Delta$ is the detuning between the bare transition frequency between electronic states and the frequency of the laser. The coefficients $V_{km}=V_0 |k-m|^{-\alpha}$ ($V_{kk} = 0$) parametrize the power-law interactions among spins in the state $\left|1\right>$, with  $V_0$ being the nearest-neighbor interaction strength and $\alpha$ the exponent with which interactions decay over distance. The terms  
$\mathcal{D}_k[\rho] =L_k \rho L_k^\dagger -\frac{1}{2}\{L^\dagger_k L_k,\rho\}$ appearing in Eq.~\eqref{eq:master_equation} are superoperators accounting for dissipative effects. Here, we consider the system to be  subject to local dephasing (at rate $\gamma > 0$), encoded in the jump operators  $L_k=\sqrt{\gamma} n_k$.

To make progress with an analytical description, it is convenient to move to a rotating reference frame, using the unitary transformation $U_t = \prod_k U_t^k$, with $U_t^k =  e^{-i \omega(t) \sigma^x_k}$, where $\omega(t) = \int_0^t \Omega(t^\prime)\df t^\prime$. 
The evolution of the transformed state $\bar{\rho}=U_t^\dagger \rho U_t$ is then governed by the master equation $\partial_t \bar{\rho} = \bar{\mathcal{L}}(t) \bar{\rho}$ with generator
\begin{equation}\label{eq:Lindblad_rotating_frame}
    \begin{split}
    \bar{\mathcal{L}}(t)[\bar{\rho}] &= -i[\bar{H}(t),\bar{\rho}] + \sum_k \bar{\mathcal{D}}_k(t)[\rho].
    \end{split}
\end{equation}
This depends on the Hamiltonian $\bar{H}(t) =\Delta \sum_kn_k(t)+\sum_{km} (V_{km}/2) n_k(t) n_m(t)$ and dissipator $\bar{\mathcal{D}}_k(t)[\rho] =  L_k(t) \bar{\rho} L^\dagger_k(t) -\frac{1}{2}\{L^\dagger_k(t) L_k(t),\bar{\rho}\}$, which contains the now time-dependent jump operators $L_k(t)=\sqrt{\gamma} n_k(t)$. 
Importantly, the generator in the rotated frame solely depends on the time-dependent projectors $n_k(t) = (U_t^k)^\dagger n_k U_t^k$, which are mutually commuting. 
This is key for the analytic construction of the $4^{N}$ %{\color{red} I added this number, because in the abstract we write "extensive". This is correct, no?} 
instantaneous (time-dependent) eigenmatrices of the Lindbladian generator, as we will show in the following: let  $|\mathbf{q}\rangle = | q_1\rangle \otimes \cdots \otimes |q_N\rangle$, where the entries of the vector $\mathbf{q}=(q_1,\ldots,q_N)$ are either $0$ or $1$. Then the rotated states $P_\mathbf{q}(t) = U_t^\dagger|\mathbf{q}\rangle\langle \mathbf{q}|U_t$ span the zero eigenspace of $\bar{\mathcal{L}}(t)$, i.e.\ the space of instantaneous stationary states.
All other eigenmatrices are of the form $ P_\mathbf{qp}(t) = U_t^\dagger|\mathbf{q}\rangle\langle \mathbf{p}|U_t$, with eigenvalues $\lambda_\mathbf{qp}= r_\mathbf{qp} + i c_\mathbf{qp} $.
The  imaginary part $c_\mathbf{qp}= \Delta \sum_k (q_k - p_k)+ \sum_{km} V_{km}(q_k q_m - p_k p_m )$  originates from the spin-spin interaction and the longitudinal field, while the real part $r_\mathbf{qp}=-\frac{\gamma}{2} \sum_k (p_k-q_k)^2$ results from the local dephasing.

\vspace{0.25cm}
\textit{Adiabatic dynamics under a single pulse.---}
We now consider the situation of a small and slowly varying transverse field, which we parametrize as $\Omega(t) = T^{-1} g(t/T)$. Here, $g:[0,1]\to \mathbb{R}$ is a smooth function with $g(0)=g(1)=0$ and $T>0$ is the total length of the pulse~[see Fig.~\ref{fig:Fig1}\bc{(b)}]. Note that the area $\int_0^T \Omega(t') \df t'$ under the pulse is then independent of $T$.
The adiabatic regime is characterized by $T\gamma\gg1$ so that one introduces also the normalized time  $s = t/T$ such that $\Omega(s)= T^{-1} g(s)$ and $\omega(s)= \int_0^s g(s') \df s'$. This takes the Lindblad master equation in the rotated frame to the standard adiabatic form
\begin{equation}\label{eq:adiabatic_master_equation}
     \frac{1}{T}\frac{\partial}{\partial s} \bar\rho(s) =  \bar{\mathcal{L}}(s)[\bar{\rho}(s)] \,,
\end{equation}
where the adiabatic parameter $T^{-1}$ enters only in front of the time-derivative. At leading order the adiabatic approximation states that the solution of Eq. \eqref{eq:adiabatic_master_equation} with initial state $\bar\rho(0) = P_{\mathbf{p}}(0)$ satisfies $\bar\rho(s) = P_{\mathbf{p}}(s) +\mathcal{O}(T^{-1})$. However, the previously derived expressions for the instantaneous spectrum allow us to analytically derive the first-order correction to the adiabatic approximation. Following Ref.~\cite[\st{Theorem 18}]{avron2012} the adiabatic density matrix (including first-order corrections) in the rotating frame is given by
\begin{equation}\label{eq:ad_expansion}
    \begin{split}
        \bar{\rho}(s)\approx& P_\mathbf{p}(s) + \frac{1}{T}\sum_{\mathbf{q}\neq \mathbf{p}} \left(\frac{P_\mathbf{q}(s) {P}^\prime_\mathbf{p}(s)}{\lambda_{\mathbf{qp}}}+\frac{{P}_\mathbf{p}^\prime (s)P_\mathbf{q}(s)}{\lambda_{\mathbf{pq}}}\right)\\
        &-\frac{1}{T}\sum_{\mathbf{q}\neq \mathbf{p}}\left(P_\mathbf{p}(s)-P_\mathbf{q}(s)\right)\int_0^s\, f_\mathbf{pq}(s^\prime) \df s^\prime\tc 
    \end{split}
\end{equation}
with ${P}^\prime_\mathbf{p}(s)=\partial {P}_\mathbf{p}(s)/\partial s$ and 
\begin{equation}
   f_\mathbf{pq}(s)=-2\frac{r_\mathbf{pq}}{|\lambda_{\mathbf{pq}}|^2}  \Tr[P_\mathbf{p}(s)({P}^\prime_\mathbf{q})^2(s)P_\mathbf{p}(s)]\geq 0.
\end{equation}
These quantities can be explicitly evaluated since the projectors $P_\mathbf{p}$ and the eigenvalues $\lambda_\mathbf{pq}$ are analytically known (see Supplemental Material \cite{supp} for details). 

At first sight, the expression in Eq.~\eqref{eq:ad_expansion} seems to be a  nonlinear one, as it involves higher powers of the projector $P(s)$. However, as we show in Ref.~\cite{supp}, this expression can be simplified and interpreted as the action of a linear superoperator  onto the initial state itself. Such a linearity allows us to explore the dynamics of initial states which are statistical mixtures of classical configurations, $\rho(0) = \sum_\mathbf{p} a_\mathbf{p}P_\mathbf{p}$.
In the original reference frame, we find that 
\begin{equation}
    {\rho}(s) \approx \rho(0) + T^{-1} \mathcal{A}(s)[\rho(0)] \approx e^{\frac{1}{T} {\mathcal{A}}(s)}[\rho(0)]  \,,
     \label{eq:ad_exponential}
\end{equation}
where the superoperator $\mathcal{A}(s)$  reads (see Ref.~\cite{supp} for details)
\begin{equation}\label{eq:ad_generator}
         \mathcal{A}(s)[{\rho}(0)] = \sum_k\left( -i[{K}_k(s),{\rho}(0)] + {\mathcal{W}}_k(s)[{\rho}(0)]\right) \, .
\end{equation}
The first contribution on the right hand side of the equality describes the coherent part of the effective dynamics. It is implemented by the local Hamiltonian terms
\begin{equation}
\label{eq:ad_Hamiltonian}
    K_k(s) = g(s) \Lambda_k \left[ \Theta_k \sigma_k^y + \frac{\gamma}{2} \sigma_k^x\right] \tc 
\end{equation}
where 
\begin{equation}
     \Lambda_k =  \Biggl[\frac{\gamma^2}{4} +\Theta_k^2 \Biggr]^{-1} \tc
\end{equation}
and $\Theta_k = \Delta + \sum_{m\neq k}V_{km} n_m$.
Here we note that the operators $\Lambda_k$ and $\Theta_k$ commute with $\sigma_k^x$ and $\sigma_k^y$, such that $K_k(s)$ is Hermitian. 
Furthermore, we observe that the first-order corrections in $T^{-1}$  generate coherences solely between classical configurations $\left|\mathbf{p}\right>$ and $\left|\mathbf{q}\right>$, which are separated by a single spin flip.
The second term on the right hand side of Eq.~\eqref{eq:ad_generator} accounts for dissipative effects through the superoperators 
\begin{equation}\label{eq:dissipative_superoperator}
    \mathcal{W}_k(s)[\rho ] = \gamma \left[ \int_0^s g^2(s^\prime) \df s^\prime \right] \Lambda_k \left(\sigma_k^x \rho \sigma_k^x - \rho \right) \tc \, 
\end{equation}
which, when acting on classical states, implement incoherent spin flips. 

The structure of the superoperator $\mathcal{A}(s)$ closely resembles that of generators which encode so-called kinetic constraints \cite{PhysRevLett.98.195702,Cancrini2008,lesanovsky2011,marcuzzi2014,PhysRevLett.126.103002,Cech_Kinetically_2025,Cech_Kinetically_2025}. These manifest in spin flip rates, which are operator valued and thus depend on the instantaneous configuration of the system. Within $\mathcal{A}(s)$ such dependence is encoded in the operator $\Lambda_k$, appearing both in the Hamiltonian and in the dissipative  dynamics as an overall prefactor [cf.~Eqs.~\eqref{eq:ad_Hamiltonian}-\eqref{eq:dissipative_superoperator}]. The presence of kinetic constraints, for instance observed in Rydberg gases~\cite{PhysRevLett.98.195702,marcuzzi2014,Bernien2017,PhysRevA.93.040701}, typically leads to an intricate slow relaxation dynamics. In the following, we show how it can affect the adiabatic evolution of our system. 

\begin{figure}[t]
    \centering
    \includegraphics[width=\linewidth,height=\textheight,keepaspectratio]{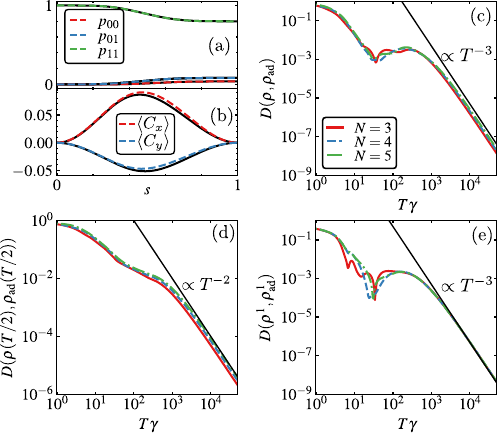}
    \caption{\textbf{Adiabatic dynamics.} (a) Populations $p_\alpha$ of classical configurations $\left|\alpha\right>$, with $\alpha=(00,01,11)$, of a system of two spins as a function of the normalized time $s$. Solid black lines are the exact dynamics and dashed lines follow from Eq.~\eqref{eq:ad_exponential}. Here, $T\gamma=400$ and $V_0 = \gamma$. The population $p_{01}$ overlaps with $p_{10}$. 
    (b) Evolution of the expectation value of the coherence operators $C_{x/y}$ (see text for definition), where the dotted lines correspond to $\rho_\mathrm{ad}$ and the black solid lines are $C_{x/y}$ computed from the exact dynamics.  (c)-(d) Trace distance, defined as $D(\sigma, \mu) = \frac{1}{2}\Tr[\sqrt{(\sigma - \mu)^\dag (\sigma - \mu)}]$, between the adiabatic approximation, Eq.~\eqref{eq:ad_exponential}, and exact dynamics as a function of $T\gamma$ for a full adiabatic pulse and a for an incomplete pulse until $t=T/2$, respectively. The number of spins is also varied, as indicated in the legend in (c). In (e), we consider the trace distance between the exact dynamics and the adiabatic approximation for the reduced density matrix of site $1$, $\rho^1 = \Tr_{2,3,\cdots}[\rho]$ for different number of spins and indicated in the legend of (c).  Other parameters: $V_0=3\gamma$, $\alpha=3, \Delta = 0$, and $g(s) = 4 \pi \sin^2(\pi s)$. The system is initially in the state $|0,0,\cdots,0\rangle$ in all simulations.}
    \label{fig:Populations_one_cycle}
\end{figure}

\vspace{0.25cm}
\textit{Numerical validation.---} In Fig.~\ref{fig:Populations_one_cycle}\bc{(a)-(b)} we show a comparison between the numerically exact evolution and the approximate evolution under Eq.~\eqref{eq:ad_exponential} for a system of two spins [more precisely, we consider the exponential application of the map $\mathcal{A}(s)$ on the initial state]. We display the populations of the classical basis states and the expectation value of the coherence  operators $C_\alpha = \sum_k \sigma_k^\alpha$, where $\alpha=x,y$. Good agreement is found for all quantities given the choice of parameters.

Next, we investigate how the error in the approximation scales as a function of the transverse field pulse length $T$ and the systems size $N$. To this end we evaluate the trace distance~\cite{nielsen2010quantum} between the state evolving under the exact dynamics and the state evolved under Eq.~\eqref{eq:ad_exponential}, both at the end and in the middle of the pulse. The results are shown in Fig.~\ref{fig:Populations_one_cycle}\bc{(c)-(d)}, respectively.
At the end of the pulse, where the approximate density matrix is diagonal [cf.~Eqs.~\eqref{eq:ad_expansion}-\eqref{eq:ad_exponential}], the error scales with $T^{-3}$. This  scaling is faster than the one expected for a first-order approximation in $1/T$. In the middle of the pulse, instead, the error shows the slower scaling $T^{-2}$, which is anyway in agreement with our approximation of Eq.~\eqref{eq:ad_exponential}. The scaling of the error with the pulse length $T$ is not affected by the interaction strength, the dephasing rate, or the number of spins, although these parameters can influence the value of $T$ necessary to enter the adiabatic regime~\cite{avron2012,sarandy2005adiabatic,PhysRevA.102.052215}.

In Fig.~\ref{fig:Populations_one_cycle}\bc{(c)-(d)}, we see how the error  depends on the system size and,  in particular, how it increases with $N$. The reason for this is that the error compares two many-body states, which  contain the full information about the system.  
On the other hand, in Fig.~\ref{fig:Populations_one_cycle}\bc{(e)}, we compute the error associated with the adiabatic approximation when looking at the reduced density matrix of a single spin, i.e., when considering reduced information on the system.  In this case, we focus on the error at the end of the pulse. While its scaling with the pulse length $T$ matches the one observed for the full density matrix, the error here does not show anymore a clear dependence on the system size. This aligns with the expectation that  adiabatic approximations for many-body systems are accurate in predicting the behavior of local observables \cite{bachmann2017adiabatic,bachmann2018adiabatictheorem,teufel2020neass}. 

\vspace{0.25cm}
\textit{Multiple pulses.---}
We will now investigate a scenario in which a number of adiabatic pulses is successively applied [Fig.~\ref{fig:Fig1}\bc{(c)}]. Each of these pulses has a shape as depicted in Fig.~\ref{fig:Fig1}\bc{(b)}, i.e. they start and end at zero transverse field strength. In the following, we will show that the ensuing many-body dynamics is in fact surprisingly well captured by an exponential map which acts on classical states, i.e. a statistical mixture of classical configurations. For a sequence of $m$ pulses this map is constructed as 
\begin{equation}\label{eq:many_pulses}
     \rho(m) \approx e^{\frac{1}{T} {\mathcal{A}}(1)} \circ \cdots  \circ e^{\frac{1}{T} {\mathcal{A}}(1)}[\rho(0)]  = e^{\frac{m}{T} {\mathcal{A}}(1)}[\rho(0)]  \tp 
\end{equation}
Here ${\mathcal{A}}(1)$ is the generator for the dynamics under a single pulse, defined in Eq. (\ref{eq:ad_generator}), evaluated at the final time ($s=1$) of the pulse. Note, that at this final time we have $K_k(1)=0$, and hence the dynamics is exclusively generated by the dissipative superoperators $\mathcal{W}_k(1)$, see Eq.~\eqref{eq:dissipative_superoperator}. This means that after the end of each pulse the state of the system remains classical, which is why the concatenation of maps leading to Eq.~\eqref{eq:many_pulses} is at all possible.

\begin{figure}[t]
    \centering
    \includegraphics[width=\linewidth,height=\textheight,keepaspectratio]{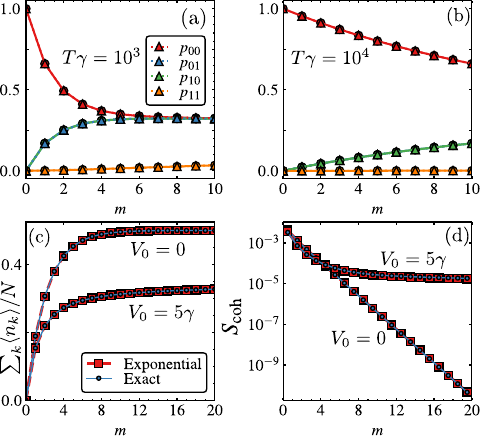}
    \caption{
    (a,b) Populations $p_\alpha$ of classical configurations $\left|\alpha\right>$, with $\alpha=(00,01,10,11)$ of a system of two spins after the application of $10$ adiabatic pulses (for $T\gamma = 10^3$ and $T\gamma = 10^4$). Triangles correspond to the exponential map given by Eq.~\eqref{eq:ad_exponential}, while circles represent the exact dynamics.  (c) Expectation value of the density of spins in the state $\left|1\right>$ during multiple pulses for the exact dynamics (blue circles) and the exponential map, Eq.~\eqref{eq:ad_exponential} (red triangles). Curves are shown for two different interaction strengths, as indicated. (d) Same as (c), but showing the entropy of coherence evaluated at the midpoint of each pulse, $s=tT^{-1}=1/2$. Other parameters: $\alpha = 3$, $\Delta = 0$, $N=4$, $T\gamma=10^3$, and $g(s) = 4\pi \sin^2(\pi s)$.} 
    \label{fig:Fig3}
\end{figure}

In Fig.~\ref{fig:Fig3}\bc{(a)-(b)} we show the comparison between the exact dynamics and the adiabatic approximation for a sequence of $10$ pulses.  
As can be seen, the state of the system evolved according to Eq.~\eqref{eq:many_pulses} is indistinguishable from the numerically exact dynamics, provided that the individual pulse length $T\gamma$ is sufficiently large. 
Note, that actually a first-order expansion of Eq.~\eqref{eq:many_pulses} in $1/T$  should in principle already yield a valid approximation for the first-order correction to the adiabatic dynamics. This would correspond to the first-order expansion in Eq.~\eqref{eq:ad_exponential}, with a function $g(s)$ that is representing a pulse train. However, this evolution would only account for single spin-flip processes. In contrast to that, the exponential map in Eq.~\eqref{eq:many_pulses} apparently correctly incorporates spin-flip processes at all orders, which is reflected in the accurate reproduction of populations, as shown by Fig.~\ref{fig:Fig3}\bc{(a,b)}.

In Fig.~\ref{fig:Fig3}\bc{(c)}, we plot the density of spins in the state $|1\rangle$, for two different values of interaction strengths and a sequence of $20$ pulses. 
We note that for $V_0=0$ the system relaxes exponentially fast to the infinite temperature state, while for $V_0=5\gamma$ the density of spins in the state $|1\rangle$ undergoes a plateau, which is a signature of slow relaxation caused by the emerging kinetic constraints~\cite{PhysRevLett.98.195702,lesanovsky2011,Cancrini2008,Cech_Kinetically_2025,marcuzzi2014,PhysRevLett.126.103002}. 

Eq.~\eqref{eq:many_pulses} in fact, even allows to develop an approximation for the state of the system at any time $t = (s + m)T$, i.e., after $m$ complete pulses plus a fraction $s$ of a pulse ($0<s<1$). The idea is that we can exploit Eq.~\eqref{eq:many_pulses} to approximate the system state after $m$ complete pulses. Since this consists of a statistical mixture of classical states, we can now use Eq.~\eqref{eq:ad_exponential} to propagate for an additional time $s$. This results in the approximation  $\rho(s + m) \approx e^{\frac{1}{T}\mathcal{A}(s)} \circ e^{\frac{m}{T}\mathcal{A}(1)}[\rho(0)]$. 
Since we are now considering times beyond single pulses, one may expect that the exact dynamics develops complex coherence patterns which are not captured by the operators $C_\alpha$. (Note indeed that these operators can only witness coherence between states differing for a single spin flip.) To be able to quantify quantum coherence in full generality and to test our approximation for $\rho(s+m)$, we focus on the entropy of coherence~\cite{PhysRevLett.113.140401,PhysRevA.93.012334}. This is defined as $S_\mathrm{coh} = S(\rho_\mathrm{diag})-S(\rho)$, where $S(\rho) = -\Tr[\rho \log \rho]$ is the von Neumann entropy and $\rho_\mathrm{diag}$ is the diagonal part of the density matrix.
Fig.~\ref{fig:Fig3}\bc{(d)} shows the entropy of coherence evaluated in the middle of a pulse ($s=1/2$) after $m$ complete pulses have already been applied, $t = (1/2+m)T$. In the figure we compare the approximation $\rho(s + m) \approx e^{\frac{1}{T}\mathcal{A}(s)} \circ e^{\frac{m}{T}\mathcal{A}(1)}[\rho(0)]$ with the exact dynamics, for both $V_0=0$ and $V_0=5\gamma$. For $V_0=0$, the system relaxes exponentially to the infinite-temperature state, resulting in an exponential decay of coherences. 
In contrast, for $V_0=5\gamma$, emergent kinetic constraints slow down the relaxation to stationarity, allowing for the state to retain quantum coherence for longer times.

\vspace{0.25cm}
\textit{Summary and outlook.---} We have considered a quantum Ising spin chain subject to local dephasing and driven via slowly varying transverse field pulses. Building on an analytic expression for the instantaneous eigenvalues and eigenstates, we have derived an explicit expression for the first-order correction to adiabatic dynamics. The ensuing effective generator has the structure of a Lindblad master equation. When the pulses start and end with a vanishing transverse field, the stroboscopic dynamics is effectively classical and governed by the presence of kinetic constraints. However, during a pulse, the evolution can also generate coherences. Our analytical results shed light into effective dynamical features of adiabatic processes in many-body systems which are currently much investigated on experimental quantum simulator platforms. In the future, it would be interesting to identify and study other classes of other adiabatically evolving dissipative systems which are amenable to an analytical treatment and which, for example, include also incoherent emission and absorption processes. It would also be interesting to further explore the approximation to the multipulse dynamics in Eq.~\eqref{eq:many_pulses} and to better understand to which order it can correctly capture adiabatic corrections. 

\begin{acknowledgements}
\noindent
We acknowledge funding from the Deutsche Forschungsgemeinschaft (DFG, German Research Foundation) under Project No. 435696605 and through the Research Units FOR 5413/1, Grant No. 465199066 and FOR 5522/1, Grant No. 499180199. This project has also received funding from the European Union’s Horizon Europe research and innovation program under Grant Agreement No. 101046968 (BRISQ). 
This work was funded by the QuantERA II Programme (project CoQuaDis, DFG Grant No. 532763411) that has received funding from the EU H2020 research and innovation programme under GA No. 101017733.
This work is also supported by the ERC grant OPEN-2QS (Grant No. 101164443, https://doi.org/10.3030/101164443).
We acknowledge the QUTIP PYTHON library~\cite{JOHANSSON20131234} used in the simulations in this work. 
\\

\textit{Data availability.---} The codes used to produce the numerical results of this work are available on GitHub~\cite{Github}.
\end{acknowledgements}

\bibliography{references}

\appendix

\newpage

\renewcommand\thesection{S\arabic{section}}
\renewcommand\theequation{S\arabic{equation}}
\renewcommand\thefigure{S\arabic{figure}}
\setcounter{equation}{0}
\setcounter{figure}{0}

\onecolumngrid

\newpage

\setcounter{page}{1}

\begin{center}
{\Large SUPPLEMENTAL MATERIAL}
\end{center}
\begin{center}
\vspace{0.8cm}
{\Large Adiabatically driven dissipative many-body quantum spin systems}
\end{center}

\begin{center}
Paulo J. Paulino$^{1,2}$, Stefan Teufel$^3$, Federico Carollo$^{4}$, and Igor Lesanovsky$^{1,5}$
\end{center}
\begin{center}
$^1${\em Institut f\"{u}r Theoretische Physik,  Universit\"{a}t T\"{u}bingen, Auf der Morgenstelle 14, 72076 T\"{u}bingen, Germany,}\\
{\em Auf der Morgenstelle 14, 72076 T\"ubingen, Germany}\\
$^2${\em  Institute for Cross-Disciplinary Physics and Complex Systems (IFISC) (UIB-CSIC),\\
E-07122 Palma de Mallorca, Spain}\\
$^3${\em Mathematisches Institut, Eberhard-Karls-Universität, Auf der Morgenstelle 10, 72076 Tübingen, Germany}\\
$^4${\em  Centre for Fluid and Complex Systems, Coventry University, Coventry, CV1 2TT, United Kingdom}\\
$^5${\em School of Physics and Astronomy and Centre for the Mathematics}\\
{\em and Theoretical Physics of Quantum Non-Equilibrium Systems,}\\
{\em  The University of Nottingham, Nottingham, NG7 2RD, United Kingdom}\\

\end{center}

\section{CALCULATION OF THE FIRST ORDER CORRECTION TO THE ADIABATIC DYNAMICS}

In this section, we report the calculation of the superoperator $\mathcal{A}(s)$ in Eq.~\eqref{eq:ad_exponential} of the main text.
The system we consider in this work,  in the original reference frame, is described  by the Hamiltonian $H(t) = \sum_{k=1}^N \left[ \Delta n_k + \Omega(t) \sigma^x_k\right] +  \frac{1}{2}\sum_{k,m=1}^N V_{km}n_k n_m $ and by the superoperators  $\mathcal{D}_k[\rho] = \gamma[n_k \rho n_k - \frac{1}{2}\{n_k, \rho\}]$, implementing local dephasing.
In order to explicitly evaluate the first order correction to the adiabatic dynamics of our system, it is convenient to  move to a rotating frame.
We do this through the unitary transformation $U_t = \prod_k U_t^k$, with $U_t^k = e^{-i\omega(t) \sigma_k^x}$, where $\omega(t) = \int_0^t \Omega(t^\prime) \df t^\prime$. 
In this frame, the density matrix, $\bar{\rho} = U_t^\dag \rho U_t$, evolves according to the master equation 
\begin{equation}
    \bar{\mathcal{L}}(t)[\bar{\rho}] = -i[\bar{H}(t), \bar{\rho}]+\sum_{k=1}^{N}\bar{\mathcal{D}}_k(t)[\bar{\rho}] \tc 
\end{equation} 
where the bar indicates that the operator is represented in the rotating frame, $\bar{O} = U_t^\dag O U_t$. 
Here, the Hamiltonian reads
\begin{equation}
    \bar{H}(t) =\Delta \sum_{k=1}^Nn_k(t)+\sum_{km} \frac{V_{km}}{2} n_k(t) n_m(t) \tc 
\end{equation} 
and the dissipative superoperators are
\begin{equation}
    \bar{\mathcal{D}}_k(t)[\rho] =  L_k(t) \bar{\rho} L^\dagger_k(t) -\frac{1}{2}\{L^\dagger_k(t) L_k(t),\bar{\rho}\} \tp 
\end{equation}
Within the rotating frame, the generator is solely written in terms of the time-dependent operators $n_k(t) = (U_t^k)^\dagger n_k U_t^k$, which mutually commute at equal time. 
We can thus obtain its spectral properties at fixed times. 
The eigenmatrices of $\bar{\mathcal{L}}(t)$ are given by $ P_\mathbf{qp}(t) = U_t^\dagger|\mathbf{q}\rangle\langle \mathbf{p}|U_t$, with eigenvalues $\lambda_\mathbf{qp}= r_\mathbf{qp} + i c_\mathbf{qp} $.
The imaginary part of the eigenvalues $c_\mathbf{qp}= \Delta \sum_k (q_k - p_k)+ \sum_{km} V_{km}(q_k q_m - p_k p_m )$  originates from the spin-spin interaction and the longitudinal field, while the real part $r_\mathbf{qp}=-\frac{\gamma}{2} \sum_k (p_k-q_k)^2$ is associated with  the local dephasing.

\subsection{First order correction to the adiabatic dynamics}

Our derivation for the first-order correction to the adiabatic dynamics of an open quantum system employs the framework established by Ref.~\cite[Theorem 18]{avron2012}.
This framework applies to systems with a Hamiltonian of the form $H = \sum_{\mathbf{p}} e_{\mathbf{p}} P_{\mathbf{p}}$, where $P_{\mathbf{p}}=|\mathbf{p}\rangle \langle \mathbf{p}|$ is a projector onto the eigenstate $|\mathbf{p}\rangle$ with eigenenergy $e_{\mathbf{p}}$.
The dissipation is given by dephasing processes, such that the  spectral properties of the dynamical generator $\mathcal{L}$ follow from the eigenvalue equation $\mathcal{L}[P_{\mathbf{pq}}] = \lambda_\mathbf{pq}P_\mathbf{pq}$, with eigenvalues $\lambda_\mathbf{pq}$ and eigenmatrices $P_\mathbf{pq}=|\mathbf{p}\rangle \langle \mathbf{q}|$.
Here, $\lambda_\mathbf{pp} = 0$, thus the eigenstates of the Hamiltonian are the stationary states of the open quantum dynamics. 
In this scenario, the first order correction to the adiabatic dynamics is given by 
\begin{equation}\label{eq:first_order_SM}
    \bar{\rho}(s)\approx  P_\mathbf{p}(s) + \frac{1}{T}\sum_{\mathbf{q}\neq \mathbf{p}} \left(\frac{P_\mathbf{q}(s) {P}^\prime_\mathbf{p}(s)}{\lambda_{\mathbf{qp}}}+\frac{{P}_\mathbf{p}^\prime (s)P_\mathbf{q}(s)}{\lambda_{\mathbf{pq}}}\right)-\frac{1}{T}\sum_{\mathbf{q}\neq \mathbf{p}}\left[P_\mathbf{p}(s)-P_\mathbf{q}(s)\right]\int_0^s\, f_\mathbf{pq}(s^\prime) \df s^\prime \tc
\end{equation}
where ${P}^\prime_\mathbf{p}(s)=\partial {P}_\mathbf{p}(s)/\partial s$, $s=t/T$ is the normalized time, and 
\begin{equation}
    f_\mathbf{pq}(s)=-2\frac{r_\mathbf{pq}}{|\lambda_{\mathbf{pq}}|^2}  \Tr[P_\mathbf{p}(s)({P}^\prime_\mathbf{q})^2(s)P_\mathbf{p}(s)]\geq 0.
\end{equation}

\subsection{Explicit evaluation of Eq.~(S4)}
Using the spectral properties of $\bar{\mathcal{L}}$, we start the explicit evaluation of Eq. \eqref{eq:ad_expansion} for our model.
We first observe that 
\begin{equation}\label{eq:projector_derivative}
    {P}^\prime_\mathbf{p}(s)=ig(s)\sum_k[ \sigma^x_k, P_\mathbf{p}(s)]
\end{equation}
and thus
\begin{equation}\label{eq:coherent_correction}
    \begin{split}
        \sum_{\mathbf{q}\neq \mathbf{p}} \left(\frac{P_\mathbf{q}(s) {P}^\prime_\mathbf{p}(s)}{\lambda_{\mathbf{qp}}}+\frac{{P}^\prime_\mathbf{p}(s)P_\mathbf{q}(s)}{\lambda_{\mathbf{pq}}}\right)  &= \sum_{\mathbf{q}\neq \mathbf{p}}\sum_k  \frac{i g(s)\langle \mathbf{q} | \sigma_k^x | \mathbf{p} \rangle   }{|\lambda_\mathbf{pq}|^2}\left( \lambda_\mathbf{qp}^* P_\mathbf{qp}(s) -  \lambda_\mathbf{qp} P_\mathbf{pq}(s) \right) \tp 
    \end{split}
\end{equation}
where we used that  $ P_\mathbf{q}(s) \sigma_k^x  P_\mathbf{p}(s) = \langle \mathbf{q} | \sigma_k^x | \mathbf{p} \rangle P_\mathbf{qp}(s)$. 
The matrix element $\langle \mathbf{q}| \sigma_k^x |\mathbf{p}\rangle$
vanishes, unless $\mathbf{p}$ and $\mathbf{q}$ differ by exactly a spin flip at site $k$.
Let $\mathbf{p}$ and $\mathbf{q}$ be such a pair of states, i.e.\ $q_k = (1-p_k)$. Then, we can write the eigenvalues, reported in the main text, as $\lambda_{\mathbf{qp}} \equiv \lambda^k_\mathbf{p} = r_{\mathbf{p}}^k + i c_{\mathbf{p}}^k$, with real part $r \equiv r_\mathbf{p}^k = -\gamma /2$ which is independent of $\mathbf{p}$ and $k$ themselves. The imaginary part is instead given by \begin{equation}
      c_\mathbf{p}^k  = \left\{ \begin{array}{cl}
         \sum_m V_{km}p_m + \Delta & \mbox{if $p_k=0$} \\
         -\sum_m V_{km}p_m - \Delta& \mbox{if $p_k=1$}.
    \end{array}
    \right.
\end{equation}
Exploiting that $\lambda_\mathbf{pq}^* = \lambda_\mathbf{qp}$, we can split Eq.~\eqref{eq:coherent_correction} into one term containing the real part of $\lambda_\mathbf{qp}$ and  another one containing  its imaginary part. 
Hence, after some algebra, we find  
\begin{equation}
    \sum_{\mathbf{q}\neq \mathbf{p}} \!\left(\frac{P_\mathbf{q}(t) {P}^\prime_\mathbf{p}(s)}{\lambda_{\mathbf{qp}}} \!+\!\frac{{P}_\mathbf{p}^\prime (s)P_\mathbf{q}(s)}{\lambda_{\mathbf{pq}}}\right) \!=\! -i\sum_k[\bar{K}_\mathbf{p}^k(s), P_\mathbf{p}(s)]. 
\end{equation}
The Hamiltonian operators
\begin{equation}
    \bar{K}_k^\mathbf{p}(s) = \frac{g(s)}{|\lambda_\mathbf{p}^k|^2} \left(c_\mathbf{p}^k\sigma_k^y(s) + \frac{\gamma}{2}\sigma_k^x(s)\right) 
\end{equation}
can generate coherence, during the adiabatic pulse, between classical configurations that are separated by a single spin flip. By assumption, these operators are zero at the beginning and the end of the pulse, $\bar{K}_k^\mathbf{p}(0) = \bar{K}_k^\mathbf{p}(1) = 0$ since $g(0)=g(1)=0$, [see Fig.~\ref{fig:Fig1}\bc{(c)} in the main text]. 

To simplify the last term in Eq.~\eqref{eq:first_order_SM}, we use, following from Eq.~\eqref{eq:projector_derivative}, that $\Tr\{P_\mathbf{p}(s)[{P}^\prime(s)]^2_\mathbf{q}(s)P_\mathbf{p}(s)\}=g^2(s)|\sum_{k} \langle \mathbf{p} |\sigma^x_k| \mathbf{q}\rangle|^2$. 
Using that $\langle \mathbf{p}|\sigma_k^x|\mathbf{q}\rangle$ is only nonzero when $\mathbf{q}$ differs only by one spin flip from $\mathbf{p}$, we find 
\begin{equation}
    \begin{split}
    \Tr\{P_\mathbf{p}(s)[{P}^\prime(s)]^2_\mathbf{q}(s)P_\mathbf{p}(s)\}&=g^2(s)|\sum_{k} \langle \mathbf{p} |\sigma^x_k| \mathbf{q}\rangle|^2\\
    &=g^2(s)\left[\delta\left(\sum_{k} (p_k-q_k)-1\right)+\delta\left(\sum_{k} (p_k-q_k)+1\right)\right] \tp 
    \end{split}
\end{equation}
The first (second) term in the second line appears when the $p_k=1(0)$ and $q_k = 0(1)$. 
Then,  we can show that 
\begin{equation}
    \begin{split}
       \sum_{\mathbf{q}\neq \mathbf{p}}\left[P_\mathbf{p}(s)\!-\!P_\mathbf{q}(s)\right]\int_0^s\!\! f_\mathbf{pq}(s^\prime) \df s^\prime &=   \frac{2r_\mathbf{pq}}{|\lambda_\mathbf{pq}|^2}\sum_{\mathbf{q}\neq \mathbf{p}}\left[P_\mathbf{p}(s)(1 - n_k)  - \sigma_k^+(s) P_\mathbf{p}(s) \sigma_k^-(s)\right]\int_0^s g^2(s^\prime) \df s^\prime \\
        &+\frac{2r_\mathbf{pq}}{|\lambda_\mathbf{pq}|^2}\sum_{\mathbf{q}\neq \mathbf{p}}\left[P_\mathbf{p}(s) n_k - \sigma_k^-(s) P_\mathbf{p}(s) \sigma_k^+(s)\right]\int_0^s g^2(s^\prime) \df s^\prime \\
        &= \frac{2r_\mathbf{pq}}{|\lambda_\mathbf{pq}|^2}\left[\int_0^s \!\! g^2(s^\prime) \df s^\prime \right]\sum_{\mathbf{q}\neq \mathbf{p}} \left[\sigma_k^-(s) P_\mathbf{p}(s) \sigma_k^+(s) + \sigma_k^+(s) P_\mathbf{p}(s) \sigma_k^-(s) - P_\mathbf{p}(s)\right]\\
        &=  \sum_k \bar{\mathcal{W}}_k^\mathbf{p}(s)[P_\mathbf{p}(s)] \tc 
    \end{split}
\end{equation}
where   
\begin{equation} 
    \bar{\mathcal{W}}_k^\mathbf{p}(s)[\circ] \;=\;  \gamma \frac{ \int_0^s g^2(s^\prime) \df s^\prime }{\frac{\gamma^2}{4} + |c_\mathbf{p}^k|^2} \left( \sigma_k^x(s) \circ \sigma_k^x(s)  - \circ\right)  \tc 
\end{equation}
takes the form of  a Lindblad superoperator.

Putting together the two contributions, from $\bar{K}^k$ and $\bar{\mathcal{W}}^k$, we find that the first-order adiabatic approximation reads
\begin{equation}
    \bar\rho(s) \approx  P_\mathbf{p}(s) + \frac{1}{T}\bar{\mathcal{A}}^{\mathbf{p}}(s) [P_\mathbf{p}(s)] 
\end{equation}
with 
$ \bar{\mathcal{A}}^{\mathbf{p}}(s)[\circ] =  -i [\bar K^\mathbf{p}(s),\circ] + \bar{\mathcal{W}}^\mathbf{p}(s)[\circ]$.

Up to now we considered the initial value problem for $\rho(0)= P_\mathbf{p}$.
By linearity we can extend the adiabatic approximation to initial states of the form $\rho(0) = \sum_\mathbf{p} a_\mathbf{p}P_\mathbf{p}$, where $a_\mathbf{p}$ is the probability to find the system in configuration given by  $P_\mathbf{p}$.
In this case, the first order correction to the adiabatic dynamics, rotated back to the Schr\"{o}dinger picture, as reported in the main text, becomes 
\begin{equation}
     {\rho}(s) \approx  \rho(0) + T^{-1}  \mathcal{A}(s)[\rho(0)] \tc 
\end{equation}
where  
\begin{equation}
    \begin{split}
         {\mathcal{A}}(s)[{\rho}(0)] = \sum_k\bigl\{ -i[{K}_k(s),{\rho}(0)] + {\mathcal{W}}_k(s)[{\rho}(0)]\bigr \} \tc 
    \end{split}
\end{equation}
with the Hamiltonian 
\begin{equation}
    K_k(s) = g(s)\Lambda_k \left(\Theta_k \sigma_k^y + \frac{\gamma}{2}\sigma_k^x \right)\tc 
\end{equation}
and the dissipator
\begin{equation}
    \mathcal{W}_k(s)[\rho ] = \gamma \left[ \int_0^s g^2(s^\prime) \df s^\prime \right] \Lambda_k \left(\sigma_k^x \rho \sigma_k^x - \rho \right) \, .
\end{equation}
Here, we have used the fact that, the quantities $c_\mathbf{p}^k$ and $|\lambda_\mathbf{p}^k|^{-2}$ that depend of the configuration $|\mathbf{p}\rangle$ can be also written as operators~\cite{lesanovsky2011} acting on the corresponding  configuration states, such that $c_\mathbf{p}^k|\mathbf{p}\rangle$ becomes  $ \Theta_k|\mathbf{p}\rangle$, where $\Theta_k= \Delta +  \sum_{m\neq k}V_{km} n_m $ and $|\lambda_\mathbf{p}^k|^{-1}|\mathbf{p}\rangle$ can be written exploiting the fact that  
\begin{equation}
     \Lambda_k |\mathbf{p}\rangle=  \Biggl[\frac{\gamma^2}{4} + \Theta_k^2 \Biggr]^{-1} |\mathbf{p}\rangle\tp
\end{equation}

\end{document}